\documentclass{iau} 
\usepackage{graphicx}

\title[Mass loss and the Eddington parameter] 
{Mass loss and the Eddington parameter}

\author[Joachim M.\ Bestenlehner]   
{Joachim M. Bestenlehner$^1$
}

\affiliation{$^1$Department of Physics \& Astronomy, Hounsfield Road, \\ University of Sheffield, S3 7RH, UK \\ email: {\tt j.m.bestenlehner@sheffield.ac.uk} 
}
\pubyear{2022}
\volume{361}  
\jname{Massive Stars Near and Far}
\editors{N. St-Louis, J. S. Vink \& J. Mackey, eds.}
\begin{document}

\maketitle

\begin{abstract}
Mass loss through stellar winds plays a dominant role in the evolution of massive stars. Very massive stars (VMSs, $> 100 M_{\odot}$) display Wolf-Rayet spectral morphologies (WNh) whilst on the main-sequence. \cite[Bestenlehner (2020)]{Bestenlehner2020} extended the elegant and widely used stellar wind theory by \cite[Castor, Abbott \& Klein (1975)]{CAK75} from the optically thin (O star) to the optically thick main-sequence (WNh) wind regime. The new mass-loss description is able to explain the empirical mass-loss dependence on the Eddington parameter and is suitable for incorporation into stellar evolution models for massive and very massive stars. The prescription can be calibrated with the transition mass-loss rate defined in \cite[Vink \& Gr\"{a}fener (2012)]{Vink2012}. Based on the stellar sample presented in \cite[Bestenlehner et al. (2014)]{Bestenlehner2014} we derive a mass-loss recipe for the Large Magellanic Cloud using the new theoretical mass-loss prescription of \cite[Bestenlehner (2020)]{Bestenlehner2020}. 
\keywords{stars: early-type, stars: massive, stars: mass loss, stars: winds, outflows, stars: atmospheres, stars: evolution}
\end{abstract}

\firstsection 
\section{Introduction}
The evolution of massive stars is dominated by mass loss and significantly influences their potential evolutionary phases (e.g. \cite[Langer 2012]{Langer2012}). In particular, stars with masses greater than 60\,$M_{\odot}$ the evolution is dominated by mass loss (\cite[Vink 2015]{Vink2015}). 
In stellar structure calculations the theoretical mass-loss descriptions by \cite[Vink et al. 2000, 2001]{Vink2000,Vink2001} are usually used for massive main-sequence stars. This mass-loss recipe under-predicts mass-loss rates for the most massive stars (e.g. \cite[Bestenlehner et al. 2014]{Bestenlehner2014}), but over-predicts at lower masses, which is also known as the weak-wind problem (e.g. \cite[Puls et al. 2008]{Puls2008}).

\begin{figure}[t]
\begin{center}
 \includegraphics[width=3.4in]{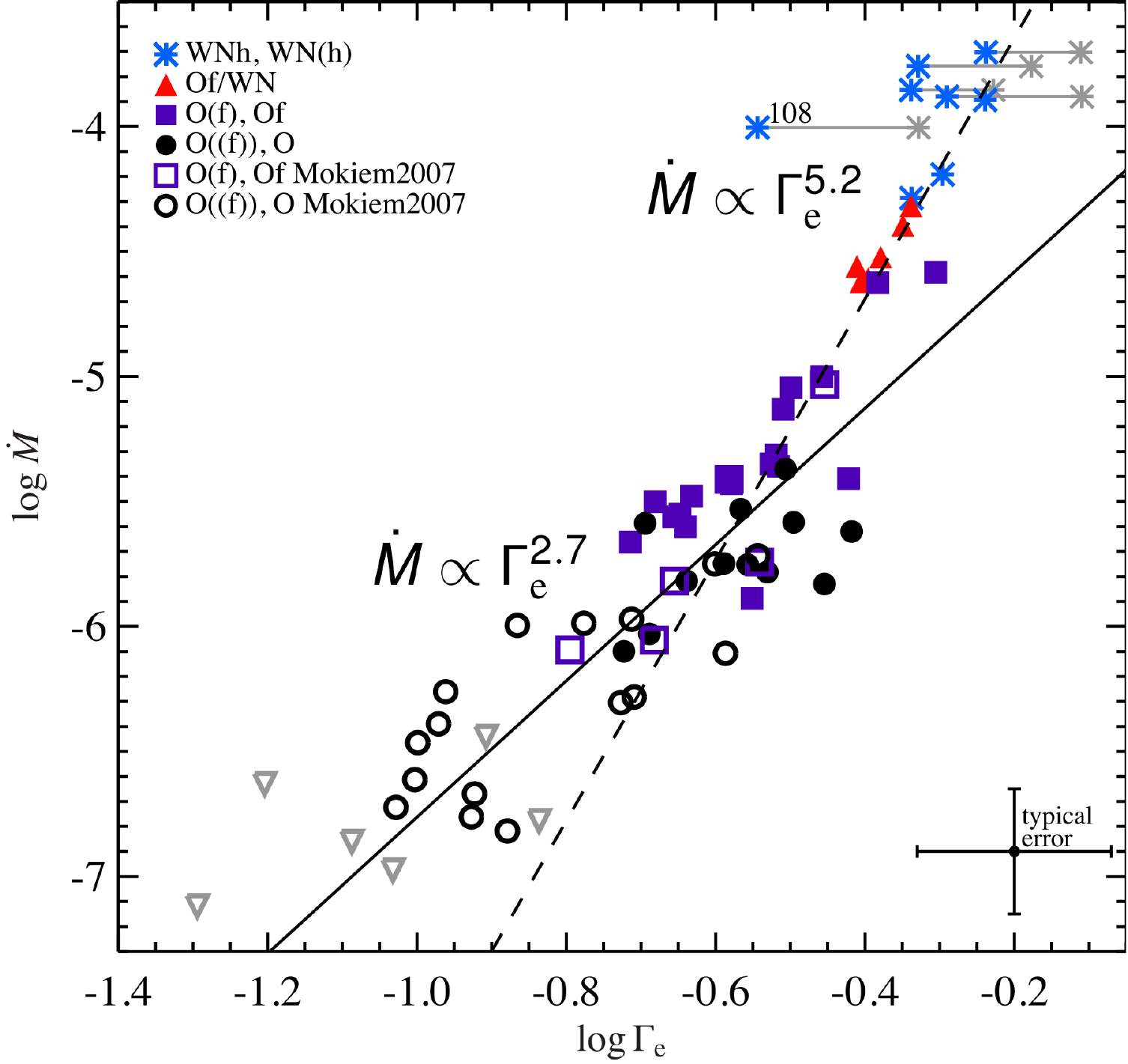} 
 \caption{$\dot{M}$ vs. $\Gamma_{\rm e}$ for several spectral types. Figure adapted from \cite[Bestenlehner et al. (2014)]{Bestenlehner2014}.}
   \label{fig1}
\end{center}
\end{figure}

Mass-loss rates depend on metallicity and the Eddington parameter $\Gamma_{\rm e}$, considering only the electron opacity, (e.g. \cite[Gr\"afener \& Hamann 2008]{Graefener2008}). \cite[Vink et al. (2011)]{Vink2011} predicted a steep increase of the mass-loss rate at the transition region from optically thin to optically thick which was empirically confirmed by \cite[Bestenlehner et al. (2014, Fig.\,\ref{fig1})]{Bestenlehner2014}.
Based on the line driven wind theory \cite[Castor, Abbott \& Klein (1975, hereafter CAK)]{CAK75} derive a mass-loss prediction for massive stars as function of stellar mass $M$, $\Gamma_{\rm e}$ and CAK parameters $k$ and $\alpha$ (see their equ. 46). However, the dependence of $\dot{M}$ on $\Gamma_{\rm e}$ was too weak to reproduce the steep increase of $\dot{M}$ when approaching the Eddington limit (Fig.~9 of \cite[Bestenlehner et al. 2014]{Bestenlehner2014}).

\section{New mass-loss description for massive and very massive stars}

The mass-luminosity relation of massive stars changes from $L\propto \mu^4 M^3$ to $L\propto M$ for $\Gamma_{\rm e} \rightarrow 1$ with luminosity $L$ and mean molecular weight $\mu$ (e.g. \cite[Yusof et al. 2013]{Yusof2013}). This behaviour is not represented in $M$ of the CAK mass-loss prediction. As shown in \cite[Bestenlehner (2020)]{Bestenlehner2020} the mass-term in the CAK mass-loss prediction can be replaced with the $M-\Gamma_{\rm e}$ relation obtained under the assumption of a fully radiative star (Eddington stellar model), which fullfills the condition $\Gamma_{\mathrm{e}} \ll 1: L\propto \mu^4 M^3$, $\Gamma_{\mathrm{e}} \rightarrow 1: L\propto M$:
\begin{equation}\label{equ1}
M \propto  \frac{1}{\mu^2}\frac{\Gamma_{\rm e}^{1/2}}{(1-\Gamma_{\rm e})^2}
\end{equation} 
with $\Gamma_{\mathrm{e}} \propto L/M$. The transition occurs at the location where $\Gamma_{\rm e}^{1/2} = (1-\Gamma_{\rm e})^2$. The transition Eddington parameter $\Gamma_{\rm e,trans,M_{Edd}} \approx 0.28$ is independent of the mean molecular weight. 
\begin{figure}[t]
\begin{center}
 \includegraphics[width=3.4in]{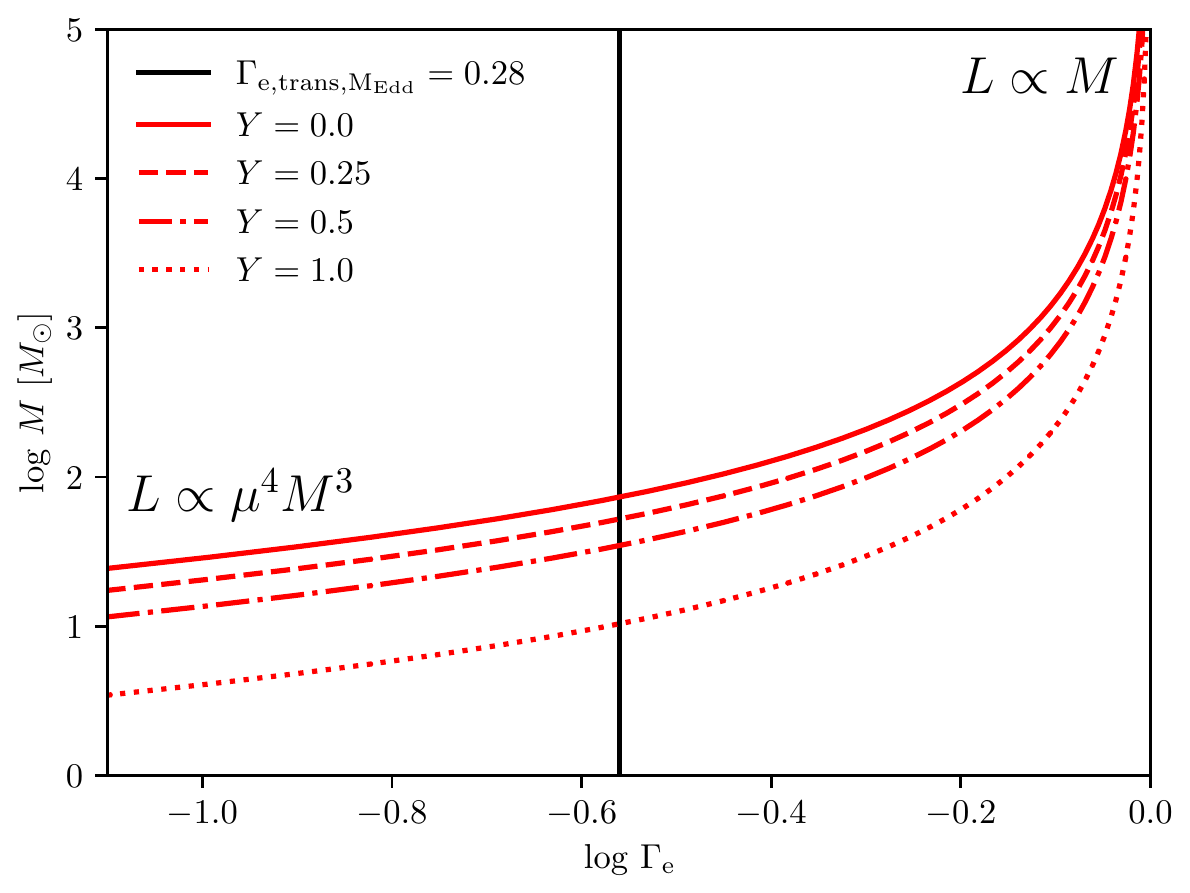} 
 \caption{Stellar mass vs. Eddington parameter for several mean molecular weights. The transition to a linear mass-luminosity relation is independent of the mean molecular weight.}
   \label{fig2}
\end{center}
\end{figure}
As shown in Fig.~\ref{fig2} the transition mass is strongly dependent on the mean molecular weight. For a pure hydrogen star the transition point occurs at $\sim75\,M_{\odot}$ while for pure helium star the transition mass is only $\sim10\,M_{\odot}$. However, stars close to the Eddington limit ($\Gamma_{\mathrm{e}} \rightarrow 1$) have hardly any dependence on $\mu$.

Substituting $M$ of the CAK mass-loss prediction with the $M-\Gamma_{\rm e}$ relation (Equation~\ref{equ1}) leads to a new CAK-type mass-loss description of the form
\begin{equation}\label{equ2}
\dot{M} \propto  \frac{1}{\mu^2}\frac{\Gamma_{\rm e}^{1/\alpha+1/2}}{(1-\Gamma_{\rm e})^{(1-\alpha)/\alpha+2}}
\end{equation}
as derived in \cite[Bestenlehner (2020)]{Bestenlehner2020}. By using the logarithmic form of Equation~\ref{equ2} and absorbing all constants into $\dot{M}_0$ we obtain
\begin{equation}\label{equ3}
\log \dot{M} = \log \dot{M}_0 + \left(\frac{1}{\alpha} + 0.5\right)\log(\Gamma_{\mathrm{e}}) - \left(\frac{1-\alpha}{\alpha} + 2\right) \log(1-\Gamma_{\rm e}).
\end{equation}
For $\Gamma_{\rm e} \ll 1$, $\dot{M} \propto \Gamma_{\rm e}^{1/\alpha+1/2}$ and $\Gamma_{\rm e} \rightarrow 1$, $\dot{M} \propto 1/(1 - \Gamma_{\rm e})^{(1-\alpha)/\alpha+2}$. At the transition, where $\left(\frac{1}{\alpha} + 0.5\right)\log(\Gamma_{\mathrm{e}}) = \left(\frac{1-\alpha}{\alpha} + 2\right) \log(1-\Gamma_{\rm e})$, we defined the transition Eddington parameter $\Gamma_{\rm e,trans}$. The CAK parameter $\alpha$ is metallicity dependent (e.g. \cite[Puls et al. 2000]{Puls2000}). The new mass-loss rate description predicts a stronger dependence on $\Gamma_{\rm e}$, larger exponent, at lower metallicity (smaller $\alpha$) and vice versa which can be observationally verified. 

\section{Discussion}
\begin{figure}[t]
\begin{center}
 \includegraphics[width=3.4in]{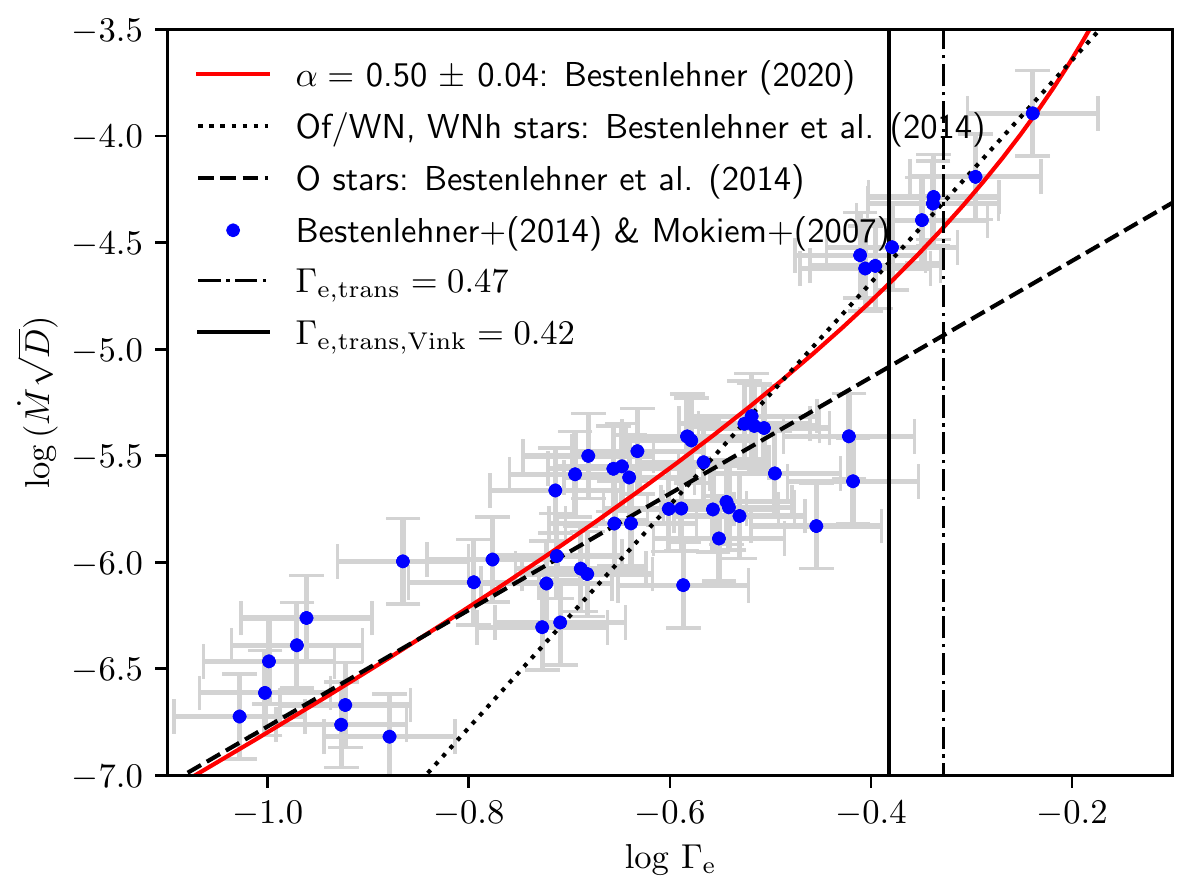} 
 \caption{Wind clumping corrected mass-loss rate vs. Eddington parameter. The derived mass-loss description by \cite[Bestenlehner (2020)]{Bestenlehner2020} well reproduce the slopes obtained by \cite[Bestenlehner et al. (2014)]{Bestenlehner2014}. The two vertical lines indicate the transitions discussed in the text.}
   \label{fig3}
\end{center}
\end{figure}
To determine the CAK parameter $\alpha$ and $\dot{M}_0$ we fitted Equation~\ref{equ3} to the data of \cite[Bestenlehner et al. (2014)]{Bestenlehner2014} and \cite[Mokiem et al. (2007)]{Mokiem2007}. The result is shown in Fig.~\ref{fig3} where we plot unclumped $\dot{M}$ against $\Gamma_{\rm e}$. 
We are able to reproduce not only the mass-loss rates of O stars but also the enhanced $\dot{M}$ of the Of/WN and WNh stars. In addition, the fit follows closely the relations derived by \cite[Bestenlehner et al. (2014)]{Bestenlehner2014} and we obtain $\alpha = 0.50 \pm 0.04$ and $\log \dot{M}_0 = -4.44 \pm 0.10$. From this it follows that $\dot{M} \propto \Gamma_{\rm e}^{2.50^{+0.17}_{-0.15}}$ for $\Gamma_{\rm e} \ll 1$, which is in good agreement with the findings by \cite[Bestenlehner et al. (2014)]{Bestenlehner2014}, $\dot{M} \propto \Gamma_{\rm e}^{2.73\pm0.43}$. The transition, where the $1/(1-\Gamma_{\rm e})^{3.00^{+0.17}_{-0.15}}$ term starts to dominate, occurs at $\Gamma_{\rm e,trans} = 0.47$ (Fig.~\ref{fig3}).

Mass-loss rates in Fig.~\ref{fig3} are unclumped under the assumption of homogeneous stellar winds. The transition mass-loss rate $\dot{M}_{\rm trans}$, introduced by \cite[Vink \& Gr\"afener (2012)]{Vink2012}, is defined at the transition from optically thin O-type star to optically thick WR star winds, where the wind efficiency $\eta$ is equal to the optical depth $\tau$ and 1, $\eta = \tau = 1$. The mass-loss rates of O and WR stars can be calibrated by utilising Equation~12 of \cite[Vink \& Gr\"afener 2012]{Vink2012}: $\dot{M} = f\,L_{\rm trans}/(v_{\infty} c) \simeq 0.6\dot{M}_{\rm trans}$ with correction factor $f\sim0.6$, transition luminosity $L_{\rm trans}$, terminal velocity $v_{\infty}$ and light speed $c$.

The transition of optically thin to optically thick winds occurs between stars of spectral type Of and Of/WN (Fig.~\ref{fig1}) with the Of/WN stars the desired transition objects (\cite[Vink \& Gr\"afener 2012]{Vink2012}). The sample of \cite[Bestenlehner et al. (2014)]{Bestenlehner2014} contains 6 Of/WN stars. Due to peculiarities we exclude VFTS\,457 (\cite[Bestenlehner et al. 2014]{Bestenlehner2014}, for more details). Based on those 5 stars a transition $\Gamma_{\rm e,trans,vink} \approx 0.42$ was determined with corresponding $\log L_{\rm trans} = 6.31\pm0.12$ and $\log \dot{M}_{\rm trans}\,[M_{\odot}/\mathrm{yr}] = -4.48 \pm 0.12$ for unclumped winds. With $L_{\rm trans}$ and the averaged terminal velocity $v_{\infty} \approx 2700 \mathrm{km/s}$ we derived an $\log \dot{M}_{\rm trans}\,[M_{\odot}/\mathrm{yr}] \approx -5.04\pm0.14$. This corresponds to a wind clumping factor $D = 13.2^{+17.0}_{-7.4}$ or volume filling factor $f_{\rm v} = 0.08^{+0.10}_{-0.04}$, which is consistent to the empirical clumping factor $D=10$ found for the Of/WN stars in \cite[Bestenlehner et al. (2014)]{Bestenlehner2014} 

On the basis of this result we calibrate $\dot{M}_0$ of Equ.~\ref{equ3} and obtain the mass-loss description for hot, massive stars in 30 Doradus in the Large Magellanic Cloud:
\begin{equation}
\log \dot{M} = -5.00 \pm 0.21~~+~~2.50^{+0.17}_{-0.15}\,\log(\Gamma_{\rm e})~~ -~~ 3.00^{+0.17}_{-0.15}\,\log(1-\Gamma_{\rm e}).
\end{equation}

This well agrees with the study of the cluster R136 in the centre of 30 Doradus by \cite[Brands et al. (2022)]{Brands2022}, who found 
$\log \dot{M} = -5.19 + 2.69\,\log(\Gamma_{\rm e})  -3.19 \,\log(1-\Gamma_{\rm e})$
on the basis of an ultraviolet+optical spectroscopic analysis.

\section*{Acknowledgements}
JMB is supported by the Science and Technology Facilities Council research grant ST/V000853/1 (PI. V. Dhillon).


\begin{discussion}

\discuss{Puls}{This approach has one big advantage, namely that they no longer
differentiate the wind features as a function of the optical depth, which more or less
always refers to the continuum and not to the lines. In the original CAK version the
question was what has the continuum to do with the different kinds of line driving for
optically thick and thin winds. As far as I have seen it here this is not the question
anymore. We have $\Gamma_{\rm e}$ from the original CAK as well as the different relation between
luminosity and mass. So this is the big advantage, but just one warning: this all assumes
that in these extreme massive O stars and main-sequence WR stars have the same mass
driving as for normal stars, so that the mass loss is not related to an iron opacity bump
or different mechanism to the CAK line driving as in the classical WR stars. If all this
would be true then all these stars (massive O and main-sequence WR stars) are just
normal and have usual O star winds. Do you agree?}

\discuss{Bestenlehner}{Yes, I agree, but there might be slightly different values of $\alpha$ at the
optically thin and thick wind regime, but if the iron-peak opacity would be the major
driver of the mass loss this parametrisation should break down. Looking at the quality of
the fit (Fig.~\ref{fig3}) there is no clear indication and an effective $\alpha$ seemed to be representative
for all those stars. Regarding classical WR stars, \cite[Sander \& Vink (2020)]{Sander2020} show that this
simple parametrisation works for some WR stars but indeed breaks down for the less
luminous ones.}

\discuss{Murphy}{Do you have a relation how $\alpha$ changes with metallicity? You say, you can
determine $\alpha$ from observations, but for very low metallicities or even zero, where you do
do not have those observations, is there a relation that you can use there?}

\discuss{Bestenlehner}{There is not an empirical relation, but \cite[Puls et al. (2000)]{Puls2000} lists theoretical $\alpha$s for several metallicities. $\alpha$ decreases with decreasing metallicity. With ULLYSES we will be able to derive empirical $\alpha$s for LMC, SMC and maybe Sextans~A metallicities as only a few OB stars are required, but this is far from zero metallicity. Once you know $\alpha$ you are able to calibrate the mass-loss descriptions with the transition mass-loss rate.} 

\discuss{Poniatowski}{You mentioned that $\alpha$ depends on metallicity, but $\alpha$ is also something that depends on the general distribution of lines, what species you have, how they are excited and so on, which can be computed, but now you can also fit that. How does this compared to general values that are normally used in literature? For instance, one typical number is $\alpha =  2/3$.}

\discuss{Bestenlehner}{The $\alpha$, you are referring to, is $\alpha'$ which is corrected for the finite cone angle, $\alpha = \alpha' - \delta$ with $\delta \approx 0.1$, e.g. \cite[Pauldrach et al. (1986)]{Pauldrach1986}. In addition, the $\alpha$ here is an effective alpha based on all types of stars and is therefore an average value. The value presented here ($\alpha = 0.5$) is still slightly lower, but derived for LMC metallicity. Using the modified CAK-theory by \cite[Pauldrach et al. (1986)]{Pauldrach1986} would have made equation~\ref{equ2} much more complex. The beauty and simplicity of this mass-loss description would have not been given anymore, even though it is sufficient to reproduce the observations.}

\end{discussion}

\end{document}